# Measuring Non-linearity in AH 2700A Capacitance Bridges with sub-ppm level uncertainty

Almazbek Imanaliev[1,2], Olivier Thévenot[1], Kamel Dougdag[1], and François Piquemal[1]

*Abstract*— The stability and non-linearity of a commercial AH 2700A capacitance bridge were studied beyond its specified capabilities using the Thompson-Lampard Calculable Capacitor (TLCC) at LNE. The TLCC allows for continuous variation of measured capacitance between 0.4 pF and 1.2 pF with a resolution of 2 parts in $10^7$ and stability better than 1 part in $10^9$ over 2 days. The study aimed to determine root cause of the saw-tooth non-linearity pattern observed in the AH 2700A capacitance bridge. This pattern becomes apparent when the internal calibration is no longer valid, indicating deviations in the bridge circuit. Additionally, the dependence of capacitance non-linearity on various factors such as frequency and capacitance value are described. This work enables automatic calibration of the commercial bridge with an uncertainty of sub-ppm level and allows for quick evaluation of TLCC's non-linearity and monitoring of any changes over time through *in-situ* measurements.

## I. INTRODUCTION

The AH 2700A commercial capacitance bridges feature a range of high performance characteristics including accuracy, stability, non-linearity and resolution [1]. Its design involves a ratio transformer and fused silica capacitance standard, which results in an accuracy of 5 ppm and non-linearity of less than 1 ppm at 1 kHz for a test load of 1 pF. These performances can be further enhanced through calibration with an appropriate capacitance standard.

To the best of our knowledge, two observations of the non-linearity of these bridges were reported in the literature [2, 3]. The first study utilizes a NIST Thompson-Lampard calculable capacitor (TLCC) with a tunability range of 0.1 pF. The study revealed that bridges of type AH 2700A with Option E (enhanced precision) exhibit exceptional nonlinear behavior [2]. Our study complement this previous research by focusing on the comprehension of the cause behind the non-linearity of the capacitance bridge employing the LNE TLCC with a tunability range of eight times greater. By having a higher tunability property, we can gain several benefits such as the ability to examine non-linear patterns over longer ranges and to obtain a more accurate estimation of the non-linearity present in the TLCC. The non-linearity patterns that we observed present distinct periodic signatures, providing a valuable tool for identifying the source of the problem. The second study performed by our group focuses on calibrating bridges using a programmable standard with very low capacitance values (10 aF to 0.1 pF) with a resolution of 10 aF, achieving an uncertainty level of 0.8 ppm [3]. In our current study, we have extended the range of capacitance values (0.4 pF to 1.2 pF), achieving a higher resolution of 0.2 aF. Using the TLCC standard has enabled us to reach an even lower level of uncertainty, with a value as low as 0.1 ppm.

This paper is an extension of the proceedings paper [4] and reports an evaluation of the non-linearity and stability of the AH 2700A bridge using the TLCC recently implemented at LNE [5]. This new standard offers a continuous variation of capacitance values within a range of 0.8 pF with a resolution of 0.2 ppm. It is highly stable due to its vacuum environment.

## II. AH 2700A CAPACITANCE BRIDGE STABILITY

Throughout this paper, the AH 2700A capacitance bridges with Option-C (continuous frequency) were employed to conduct precise and accurate 3-terminal measurements of capacitance standards at audio frequency ranges. The detailed specifications for these bridges are outlined in Table 1 for the experimental parameters used in this study. The information on the specifications was obtained utilizing the excel spreadsheet provided by the manufacturer through its website [1]. In the measurement process, a bridge voltage of 15 V was applied and the dissipation factor was around 0.0001, unless otherwise stated.

The primary focus of this study was on the non-linearity, resolution, and stability of the capacitance bridge. As per the data presented in Table 1, it is expected that these parameters will be equal or less than 1 ppm in a typical measurement cycle. To ensure the validity of the results, all measurements were performed in a controlled environment, regulated to maintain a temperature of $20 \pm 0.3$ °C and a relative humidity of $50 \pm 10$ %.

TABLE I
AH 2700A PERFORMANCE SPECIFICATIONS

| Set parameters in this paper | | | Expected performance | | | | |
|---|---|---|---|---|---|---|---|
| Load capacitance | Averaging time | Measurement frequency | Resolution | Non-linearity | Accuracy | Stability | Temperature coefficient |
| pF | s | kHz | ppm | ±ppm | ±ppm | ±ppm/year | ±ppm/°C |
| 1 | 6.9 | 1 | 0.9 | 0.7 | 5.9 | 1.2 | 0.77 |
| 1 | 4.9 | 2 | 1 | 0.8 | 7.1 | 1.4 | 0.95 |
| 10 | 6.9 | 1 | 0.2 | 0.2 | 5.2 | 1 | 0.1 |

[1]Authors are with the Laboratoire national de métrologie et d'essais (LNE), 29 avenue Roger Hennequin, Trappes, FRANCE

[2]Author to whom correspondence should be addressed (almazbek.imanaliev@lne.fr)

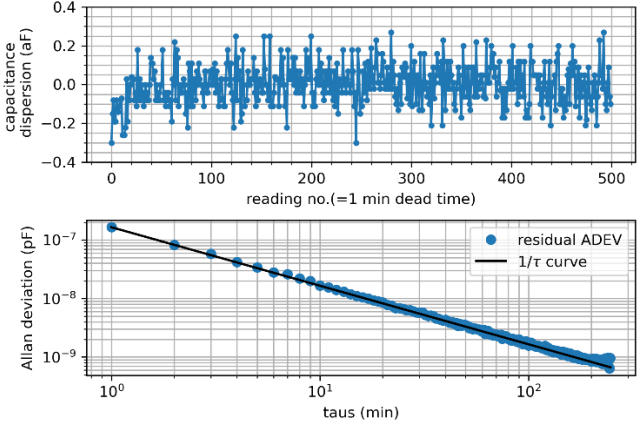

**Fig. 1.** Capacitance of a AH 11A standard measured with the AH 2700A at 1 kHz for 8 hours. (top) Dispersion of the measurement around the mean value of 1 pF, (bottom) Allan deviation of the dispersion in blue and curve with slope = -1 (white phase noise) in black.

This ensured that variations in temperature and humidity would not affect the measurements.

Figure 1 illustrates an 8-hour continuous capacitance measurement of an AH 11A fused-silica capacitance standard using the capacitance bridge. The reference AH 11A capacitor, known for its exceptional stability, was used to analyze the noise and stability of the AH 2700A capacitance bridge to its limits of specifications. As shown in figure 1, the peak-to-peak dispersion of measurements was found in the order of 3 parts in $10^7$, and the stability was found to be at an outstanding level of 1 part in $10^9$, even after more than three hours of measurements.

## II. EXPERIMENTAL SETUP

The experimental setup used to measure the capacitance of the TLCC with a AH2700A bridge is depicted in figure 2. For the purpose of this paper, only the mean cross-capacitance of the TLCC is considered:

$$C_m(z) = \frac{C_{1,3\oplus 4}+C_{2,4\oplus 5}+C_{3,5\oplus 1}+C_{4,1\oplus 2}+C_{5,2\oplus 3}}{5} \quad (1)$$

where $C_{i,j\oplus k}$ is the cross-capacitance between electrode $i$ and opposite electrodes $j,k$. For instance, $C_{1,3\oplus 4}$ refers to the measured capacitance between the electrode 1 and the electrodes 3 and 4, which are electrically linked.

The multi-switch in the setup allows us to permute between five possible combinations of $C_{i,j\oplus k}$. The mean cross-capacitance $C_m(z)$ varies in a linear manner as a function of the upward or downward movement (vertical displacement) $z$ of the mobile electrode. This movement is precisely measured using a Michelson interferometer. The movable electrode in the center of the five electrodes (figure 2) is displaced in a vertical direction with a step of approximately 56 nm by using a DC brushless motor equipped with an integrated Hall sensor.

The lateral position of the movable electrode is precisely restored for each vertical displacement by means of piezoelectric actuators and by measuring the capacitance between the movable electrode and the neighboring electrodes. The entire system placed in vacuum is piloted by a personal computer, which manages the drive motor, the piezoelectric actuators, the multi-switch, and records the capacitance and displacement values in a measurement sequence.

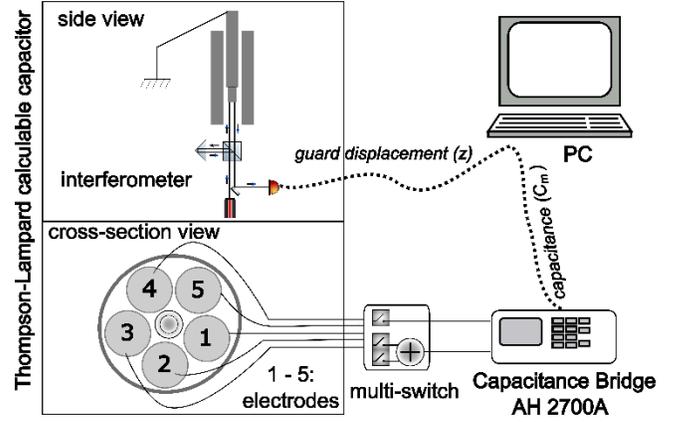

**Fig. 2.** Measurement setup

## III. THE SETUP STABILITY

Initially, the stability of the entire experimental setup measurement chain outlined in the previous section was evaluated. To achieve this, the position of the movable guard was adjusted such that $C_m \approx 1$ pF. Subsequently, the cross-capacitance measurement was taken every minute for a duration of 2 days.

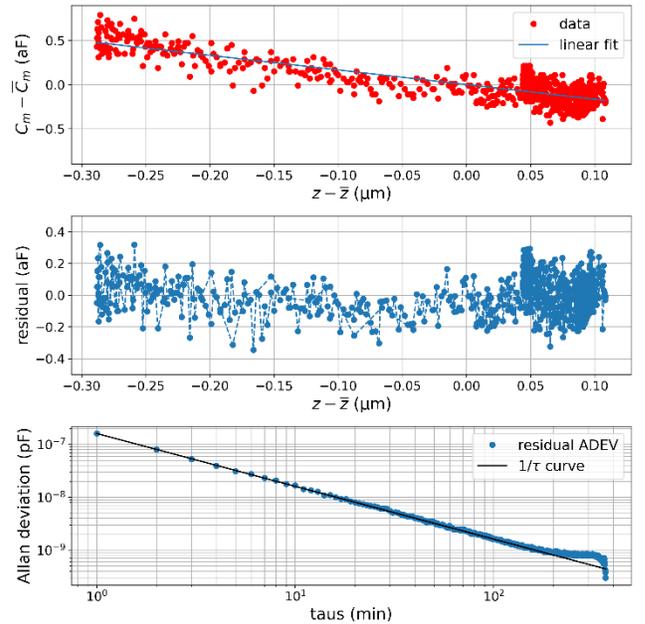

**Fig. 3.** (top) The mean cross-capacitance $C_m$ deviating from its mean value is monitored for 2 days. (middle) Residual after adjusting for changes in $z$. (bottom) Allan deviation of the residual dispersion in blue and curve with slope = -1 (white phase noise) in black.

This measurement is similar to the measurement presented in figure 1, with the exception that the reference standard is replaced by the TLCC. The outcome of the measurement is represented in figure 3.

The position of the movable guard is not locked, and over time, it may deviate from its original position. This can be observed on the top plot of figure 3, where a micrometer-level change in the movable guard position is apparent. As expected, there is a linear correlation between $C_m(z)$ and the interferometer reading $z$ as can be observed from the results shown in figure 3. By taking into account the deviation of the movable guard electrode's position, the residual is obtained in the middle plot.

The residual is found to exhibit the same peak-to-peak dispersion of 3 parts in $10^7$ as that previously observed with the AH 11A capacitance standard (figure 1). Same results have been obtained on the stability of the residual, with a level of 1 part in $10^9$ for a same period of time.

### IV. NON-LINEARITY OF AH 2700A

*A. Before internal calibration of AH 2700A*

To investigate the non-linearity of the AH 2700A capacitance bridge, the mean capacitance of the TLCC standard is varied around its nominal value of 1 pF by displacing the movable electrode by approximately 5 mm with a step size of 0.2 mm. The residual of the linear fit on $C_m(z)$ provides insights into the non-linearity of the bridge. Figure 4 illustrates the obtained residuals for different parameters of the measurement.

The top graph of figure 4 displays the best-fit residuals of three different units (AH20, AH28, AH22) of the same AH 2700A model. A saw-tooth pattern can be easily observed for each bridge with distinct periodical behavior.

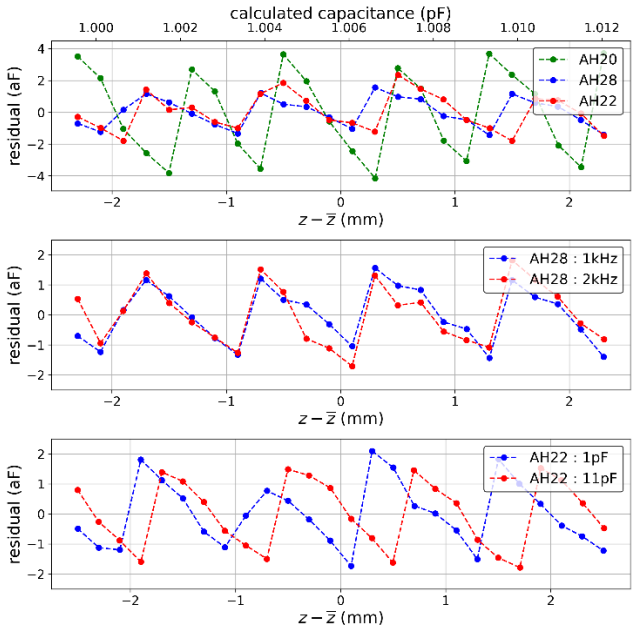

**Fig. 4.** Non-linearity of AH2700A capacitance bridges (top) as a function of frequency (middle) and capacitance value (bottom).

The amplitude of this pattern can vary among the bridges, reaching up to ± 4 ppm. This behavior is substantially different from the expected non-linearity level according to the specifications listed in Table 1.

The middle graph of figure 4 compares two measurements taken with the same bridge at two frequencies 1 kHz and 2 kHz. These results are comparable, indicating that the saw-tooth patterns do not depend on the frequency of the instrument.

The bottom graph of figure 4 shows the impact of changing the measured capacitance value from 1 pF to 11 pF by connecting a 10 pF standard in parallel to the TLCC standard. The amplitude of the saw-tooth pattern remains unchanged, but a phase shift is observed, which is expected as the added standard does not have an exact multiple of the 1 pF TLCC standard.

One may question whether these saw-tooth patterns can be attributed to mechanical imperfections on the TLCC electrode surfaces. However, this is highly unlikely as the mechanical imperfections would tend to be of a long-range nature due to the manufacturing process. Additionally, it is clearly evident that the same pattern is present in all five $C_{i,j \oplus k}$ measurements, as shown in figure 5. It is improbable to reproduce the same defect unintentionally with such perfect correlation on each electrode surface.

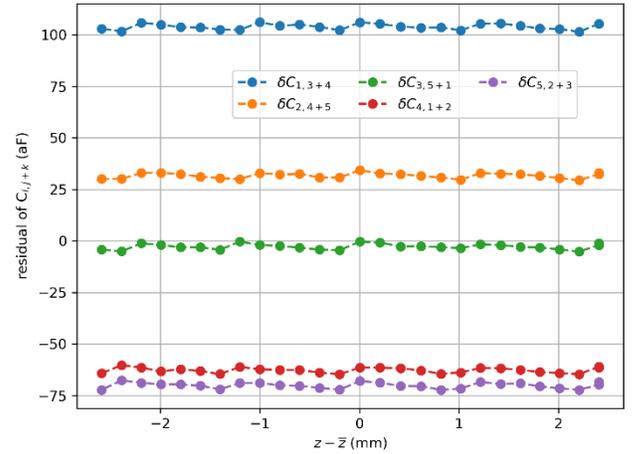

**Fig. 5.** Saw-tooth patterns in all $C_{i,j \oplus k}$ are identical, indicating non-linearity in the capacitance bridge.

To better understand the cause of the saw-tooth non-linearity pattern, a thorough examination of the AH 2700A bridge circuit design is required. The AH 2700A bridge contains a true bridge in the conventional sense, comprising a ratio transformer and a set of fused-silica reference capacitors. Both the resistive and capacitive components of an unknown capacitance $C_x$ are independently balanced. Our focus will be solely on the in-phase component for the following analysis. The AH 2700A bridge balance equation can be expressed in simpler form as follows:

$$\omega u r_x C_x = \omega u [r_1 C_A + r_2 C_B + r_{3-8} C_C] \quad (2)$$

where $C_x = C_m(z)$ since the DUT is the TLCC standard, $\omega$ and $u$ are the angular frequency and the voltage respectively at which the ratio transformer is excited using a sine wave generator. $C_A$, $C_B$ and $C_C$ are the capacitances of the temperature controlled fused-silica reference capacitors with nominal values related as $C_A = 10C_B = 10C_C$. $r_x, r_1, r_2$ and $r_{3-8}$ are transformer ratios which switch the driving voltages of corresponding reference capacitors and are adjusted independently by selecting the transformer taps during a bridge balancing procedure. There are 12 possible taps from $10.0u$ to $-1.0u$ for every ratio selection.

The bridge balance is adjustable with precision of up to eight decades, i.e. eight decimal places. Adjustment of the two most significant decades, corresponding to the initial two terms of the right-hand side of (2), is directly achieved by selecting the main ratio transformer taps $r_1$ and $r_2$ with reed relays.

To adjust the remaining decades (third to eighth decade), the bridge uses the ratio transformer multiplying D to A converter (RTMDAC) in combination with an operational amplifier to drive the reference capacitor $C_C$. The operational amplifier functions as a precision adder by incorporating input $R_i$ and feedback $R_f$ resistors. The output voltage $ur_{3-8}$ is the sum of each decade input voltage $ur_i$ multiplied by the feedback resistance and divided by the respective input resistance: $ur_{3-8} = u\sum_{i=3}^{8} r_i \frac{R_f}{R_i}$. This configuration, combined with the driven reference capacitor $C_C$, makes the precision adder and its associated input and feedback resistors functionally equivalent to a set of simulated reference capacitors $C_i$. The value of each full decade's equivalent capacitor is determined by multiplying the value of the driven reference capacitor by the ratio of the feedback resistor to the corresponding input resistor value: $C_i = C_C \frac{R_f}{R_i}$. So, the equation (2) can be rewritten as:

$$\omega u r_x C_m(z) = \omega u [r_1 C_A + r_2 C_B + \sum_{i=3}^{8} r_i C_i] \quad (3)$$

where voltage ratios $r_i$ are any integer from -1 to 10 and are adjusted independently by selecting ratio transformer taps on every input resistor of the adder. This task is fulfilled by the RTMDAC.

The AH 2700A bridge's internal calibration function involves comparing a decade with a ratio of $r_i = -1$ to its adjacent decade of lesser significance, which has a ratio of $r_{i-1} = 10$.

Ideally the sum of the results should be zero: $r_i C_i + r_{i-1} C_{i-1} = 0$. If the sum is non-zero, the result is amplified and measured to correct the decade of lesser significance. This procedure allows calibration of all the decades relative to the most significant one.

However, over time, the values of the associated resistances may vary, rendering the internal calibration invalid. The observed saw-tooth non-linearity pattern can be attributed to this effect. In this scenario, the staircase effect seen in figure 4 is a direct measurement of a discrepancy between adjacent decades' equivalent capacitors: $\varepsilon = C_i - 10C_{i-1}$.

To provide further evidence for our explanation, we conducted a similar measurement as in figure 4, but with a wider range of capacitance variation. We present the results in figure 6, which reveal a staircase effect of the next more significant decade: $\varepsilon' = C_{i+1} - 10C_i$ in addition to the one $\varepsilon$ previously observed. It is worth noting that there are nine stairs ($\varepsilon$) preceding the more significant stair ($\varepsilon'$). This is a clear signature of transformer ratio increments $r_i$ from 0 to 9. Increments of the less significant decade ratio $r_{i-1}$ are the linear part of smaller stairs ($\varepsilon$) as can be seen in figure 6.

The non-linearity of the bridge is measured immediately after applying the internal calibration procedure, and the results are presented in the following section.

### B. After internal calibration of AH 2700A

The AH 2700A capacitance bridge features highly stable reference capacitors and ratio transformers. However, other components within the bridge may be less stable, and may drift over time or with changes in temperature. To ensure accuracy and consistency in measurements, an internal calibration process is performed, which automatically verifies the performance of these components relative to the reference capacitors and ratio transformers. This internal calibration process corrects for any changes that may occur over time or due to differences in ambient temperature.

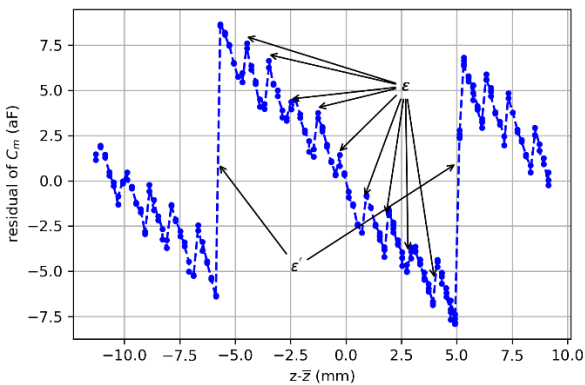

**Fig. 6.** Step errors originating from input resistors of three different decades (see text for details).

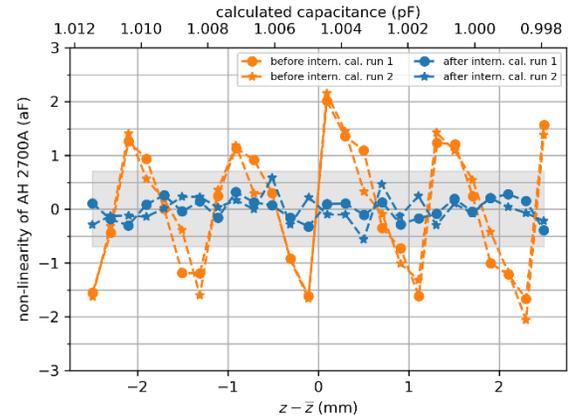

**Fig. 7.** Capacitance bridge non-linearity before and after internal calibration. Gray region represents the non-linearity specified in the Table 1.

Without proper calibration, small offsets, steps, or staircase (saw-tooth) effects may be present in the measurements as a function of the actual capacitance or loss tangent value as it was observed in the last section. These errors are too complex to predict or correct externally, making it crucial to perform frequent internal calibrations to minimize them. According to the user's manual of the instrument, internal calibrations, the procedure described on page 9-4 of the operating manual [1], should be performed every month or two for maximum precision, even if the ambient temperature is constant.

The internal calibration mainly corrects for errors in the linearity of the AH 2700A. The measurements from the previous section were repeated before and after an internal calibration of one of the bridges. The results are shown in figure 7. It is clear that the calibration completely corrected non-linearity errors and brought the apparatus in line with its specifications.

## IV. NON-LINEARITY OF THE TLCC STANDARD

We have previously varied the capacitance of the TLCC within a limited range around 1 pF, but now we plan to explore the full range of 0.4 pF to 1.2 pF and conduct similar analyses. Our focus will be on the non-linearity of the TLCC, which can greatly impact uncertainty of the standard and should be minimized to 1 part in $10^8$ level [5, 6]. The non-linearity is closely tied to the quality of the coaxiality of the TLCC's electrodes, which can be affected by factors such as conicity, surface defects, and electrode alignment. Our current setup allows for quick evaluation of the TLCC's non-linearity and monitoring of any changes over time through *in-situ* measurements.

In figure 8, the mean capacitance $C_m(z)$ of a TLCC standard for a single measurement run is plotted as a function of the movable guard's displacement $z$. The full range of displacement is more than 300 mm, which resulted in a variation of $C_m(z)$ by more than 0.8 pF.

The slope of this curve is an estimation of the mean capacitance per unit length, whose theoretical value is

$$2\gamma = 2\frac{\epsilon_0}{\pi}\ln\frac{2}{\sqrt{5}-1} = 2.71247125\ldots \text{ pF m}^{-1}$$

for a perfectly symmetrical five-bar standard [6]. The linear fit

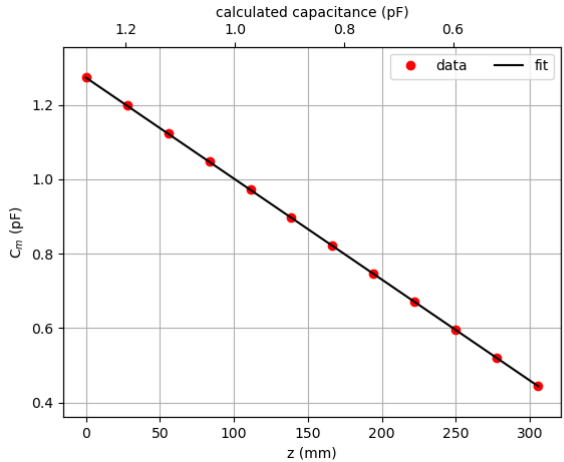

**Fig. 8.** The linear relationship between the TLCC mean capacitance $C_m(z)$ and the movable electrode's displacement $z$.

The calculated capacitance is derived from the theoretical capacitance per unit length and the measured $z$ values.
of the data in figure 8 yields a mean capacitance per unit length of 2.7124723 pF m$^{-1}$ with a standard deviation of $3.5 \cdot 10^{-7}$ pF m$^{-1}$. This value is in good agreement with the theoretical value of 2.7124725 pF m$^{-1}$, considering the accuracy of the AH 2700A bridge used in the measurement. Furthermore, the difference between the two values indicates the level of symmetry present in the TLCC electrode system.

In the next step, we examine the residual of the data shown in figure 8 after it has been adjusted using linear regression. As previously stated, this residual directly quantifies the non-linearities present in either the AH 2700A bridge or the TLCC standard, whichever has the stronger influence. The resulting residual is illustrated in figure 9, with and without internal calibration of the AH 2700A capacitance bridge. The inset plot of figure 9 is the same curve, which was presented in figure 7. For full range capacitance variation, we observe the same level of non-linearity of AH 2700A capacitance bridge as previously seen for short-range capacitance variation. This is a clear indication that the TLCC non-linearity is less than 5 parts in $10^7$, as was expected based on the evaluation of the alignment of the electrodes. In the near future, we plan to carry out a more comprehensive study of the TLCC non-linearity at an extremely precise level of one part in $10^8$ by using a specialized metrological capacitance bridge.

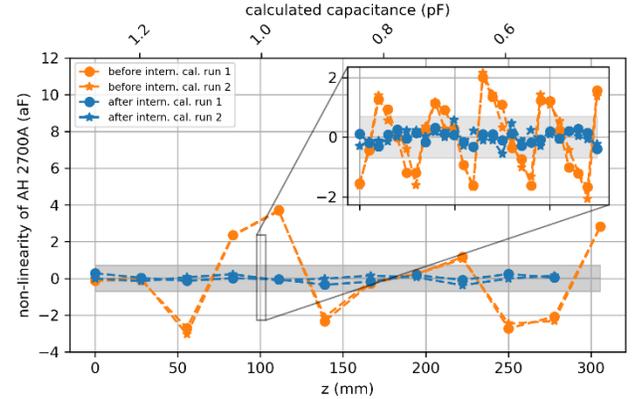

**Fig. 9.** The residual of the TLCC capacitance full range variation before and after internal calibration is depicted in blue and orange respectively. The non-linearity of the AH 2700A is the principal factor of this residual level. The gray region in the graph represents the non-linearity specified in Table 1.

## V. CONCLUSION

In conclusion, the study demonstrated the effectiveness of the LNE Thompson-Lampard calculable capacitor standard and its automated setup in evaluating the stability and non-linearity of the AH2700A capacitance bridge. The TLCC standard's extended tunability range of 0.8 pF and fine resolution of 0.2 aF make it an excellent tool for studying non-linearity. Our findings revealed that resistances' variations over time in the capacitance bridge circuit were the primary source of observed non-linearity and highlighted the significance of internal calibration when operating the AH2700A bridge. Additionally, we investigated how capacitance non-linearity is affected by factors such as frequency and capacitance value. Our work enables automatic calibration of the commercial bridge with an impressive uncertainty of sub-ppm level.

Importantly, our study demonstrated the effectiveness of using the Thompson-Lampard standard as a variable reference for evaluating non-linearity in the bridge. This methodology can be adopted for various applications, including evaluating non-linearity in other variable electric standards, such as voltage synthesizers, digitizers, impedance meters, variable inductors, capacitors, and resistors. By adopting this approach, the uncertainty of non-linearity checks can be improved, leading to more accurate measurements. Our future work will build upon these findings to further refine this methodology and explore its potential in a broader range of applications.

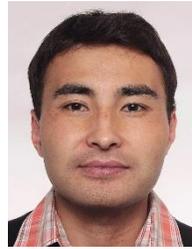
**Almazbek Imanaliev** was born in Naryn, Kyrgyzstan, in March 1988. He received the PhD degree in quantum atom optics from the University of Paris-Saclay ( Laboratory Charles Fabry) in 2016. His activity was first focused on the cold atom gravimeter with the LNE-SYRTE. Since 2019, he is a research engineer at LNE in the fundamental electrical metrology department and involved in the development of Thompson-Lampard Calculable Capacitor and the digital impedance bridges.

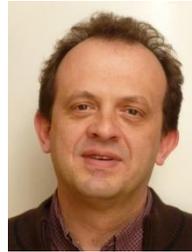
**Olivier Thevenot** was born in Paris in 1968. He received the engineer degree in "Measurements and Instrumentation" from the CNAM, Paris in 2001. He is currently project manager in electrical Metrology. He has been working for more than 20 years in the field of low frequency electrical metrology. He is mainly involved in the development of the French national standards of capacitance and resistance.

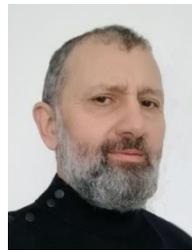
**Kamel Dougdag** was born in Algiers, Algeria, in September 1965. He obtained the DUT "Mechanical and Production Engineering at the University of Paris 8" in 1993. He is currently a research engineer in the Kibble Balance project team. He has worked for more than 30 years in the field of mechanical design. He is involved in mechanical developments within the department of fundamental electrical metrology. He participates in the mechanical development of Thompson-Lampard Calculable Capacitor, the Kibble balance, Milligram mass, quantum Hall effect equipment (QHE) and Scanning Microwave Microscope (SMM) equipment.

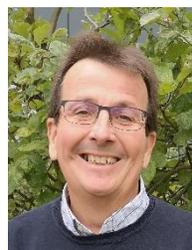
**Francois Piquemal** was born in Bois d'Arcy, France, in October 1960. He received the PhD degree in condensed matter physics from the Université de Jussieu, Paris (1988) and the habilitation degree (accreditation to supervise research) in Sciences for Engineer from the Ecole Normale Supérieure ENS-Cachan (2013). Since 1988, he has been with the French National Metrology Institute implied in the domain of Electricity and Magnetism, presently LNE (formerly LCIE). His activity was firstly focused on the quantum Hall effect (QHE) and the resistance metrology (including CCC based resistance bridge). Since 2001, he was in charge of the fundamental electrical metrology department whose activities are dedicated to quantum metrology, SI realizations of the units and determination of the fundamental constants (calculable capacitor, Kibble balance …) and more recently include electrical nanometrology.